# Self-Interference Cancellation Using Time-Domain Phase Noise Estimation in OFDM Full-Duplex Systems

Heba Shehata and Tamer Khattab

### Abstract

In full-duplex systems, oscillator phase noise (PN) problem is considered the bottleneck challenge that may face the self-interference cancellation (SIC) stage especially when orthogonal frequency division multiplexing (OFDM) transmission scheme is deployed. Phase noise degrades the SIC performance significantly, if not mitigated before or during the SIC technique. The presence of the oscillator phase noise has different impacts on the transmitted data symbol like common phase error (CPE) and inter-carrier interference (ICI). However, phase noise can be estimated and mitigated digitally in either time or frequency domain. Through this work, we propose a novel and simple time domain self-interference (SI) phase noise estimation and mitigation technique. The proposed algorithm is inspired from Wiener filtering in time domain. Simulation results show that the proposed algorithm has a superior performance than the already-existing time-domain or frequency domain PN mitigation solutions with a noticeable reduction in the computational complexity.

### Index Terms

Self-interference cancellation; OFDM; full-duplex; phase noise; time-domain estimation; Wiener filter.

## I. INTRODUCTION

Although, full duplexing is an old communication trend it was infeasible till recently. Full-duplex (FD) systems utilize all the communication resources and save all the degrees-of-freedom of the two-ways transmission scheme efficiently. Self-interference (SI) is the main problem that limited the FD system feasibility till recent revolutionary progress achieved in digital signal processing (DSP)

Heba Shehata[*] and Tamer Khattab[†] are with the Department of Electrical Engineering at Qatar University, Doha, Qatar. (e-mail: h.m.shehata@ieee.org[*], tkhattab@ieee.org[†])

techniques [1]. The problem is that a strong self-jamming signal interferes with the received signal with a much higher power level than the signal of interest (SOI) [2]. It is required to suppress up to 90-110dB of interference power to ensure that it is below the noise floor of the receiver. Different self-interference cancellation (SIC) techniques were developed and used in FD systems [4-9].

It was proven that phase noise problem is one of the most challenging factors that face the self-interference cancellation stage in FD systems as phase noise spread degrades the performance of the SIC technique significantly [3] especially with the case of OFDM transmission. In OFDM transmission schemes, phase-noise from oscillator instabilities in both the transmitter and receiver is a potentially serious problem, especially when bandwidth efficient, high order signal constellations are employed.

The two effects of phase-noise on the transmitted symbol are: the common phase error (CPE) and inter-carrier interference (ICI). These two effects were considered when the impact of the PN on OFDM systems was studied by different literatures [4-6,9]. Yet, viewing the PN impact on the OFDM symbol from another perspective, ICI can be analyzed as loss of orthogonality and subcarrier-wise phase noise spread. Subcarrier-wise PN spread affects the value of phase error imposed on a certain data symbol [7]. In fact, studying the phase noise from this point of view adds some complexity to its estimation given ICI or CPE in the frequency domain.

Different approaches have been proposed to eliminate PN effects in OFDM systems [4-6,9]. These approaches are categorized as time-domain approaches [6,8], and frequency-domain approaches [4,5,8]. The time-domain approaches mitigate phase noise at the receiver before the digital Fourier transform (DFT) and deal with the received signal samples. The frequency-domain approaches perform the correction of CPE and ICI to a certain precision in the frequency domain.

In this paper, we propose a new simple time-domain PN estimation approach based on Wiener filtering algorithm that achieves a high SI suppression level with a superior PN mitigation results when compared with two current SIC techniques in the presence of phase noise [6,8]. The proposed algorithm shows a high SI suppression level with a clear reduction in the computational complexity. This paper is organized as follows: System model with the PN effect on the transmitted signal and recent SIC techniques in the presence of PN are briefly reviewed in Section II. Section III introduces the proposed time-domain PN estimator based on the optimal Wiener filtering. Simulation results and analysis are shown in Section IV and Section V presents the conclusion.

II. SYSTEM MODEL

In OFDM Full duplex transmission, transceivers may suffer from high self-interfering signal from its own transmitter which shares the used frequency band and time slot with its received signal of interest (SOI). For this case, many approaches were studied to reduce this power significantly or to cancel it if possible either by using passive cancellation [10], analog SIC[11], digital SIC [12] or combining different approaches. The block diagram of the OFDM FD transceiver is shown in Fig. 1.

The used full-duplex transceiver model is simplified, so that the nonlinear amplification and filtering stages or any system impairments are left out and the effect of the phase noise is studied separately. Starting from the transmitter, the OFDM samples to be transmitted are converted to an analog signal waveform using the digital-to-analog converter (DAC), and then upconverted to the carrier frequency. This stage adds the PN impairment to the transmitted signal in the time domain.

At the receiver with a proper antenna separation [10], the RF signal is fed to the analog self-interference cancellation which effectively removes most of the SI signal from the main multipath component of the channel. The received baseband time-domain signal before the digital SIC can be written as [5]

$$y_n = \left[\left(x_n^I e^{j\phi_n^{t,I}} * h_n^I\right) + \left(x_n^S e^{j\phi_n^{t,S}} * h_n^S\right)\right] e^{j\phi_n^r} + z_n \qquad (1)$$

Where $n$ is the time sample index, $x^I$, $x^S$ are the residual SI signal after RF and analog SIC and the signal-of-interest (SOI), respectively. $\varphi^{t,I}$ and $\varphi^{t,S}$ are the SI and SOI transmitter PN and $\varphi^r$ is the receiver PN, $h^I$, $h^S$ are SI and SOI channels and $z$ is the added AWGN.

The residual SI is then suppressed using the digital SIC stage after the analog-to-digital converter either in time or frequency domain. The problem with the SIC technique is that the phase noise is considered a random process and degrades the performance of the SIC technique effectively if not compensated. Taking DFT of (1), the phase noise impact in the frequency domain can be represented as

$$Y_k = \sum_{m=0}^{N-1}\sum_{l=0}^{N-1} X_l^I H_m^I J_{m-l}^{t,I} J_{k-m}^r + \sum_{m=0}^{N-1}\sum_{l=0}^{N-1} X_l^S H_m^S J_{m-l}^{t,S} J_{k-m}^r + Z_k \qquad (2)$$
$$= Y_k^I + Y_k^S + Z_k$$

where $k$ is the OFDM subcarrier index, $N$ is the total number of OFDM subcarriers, $Y_k^I, Y_k^S$ and $Z_k$ are the frequency-domain representation of the SI, SOI and the AWGN noise.

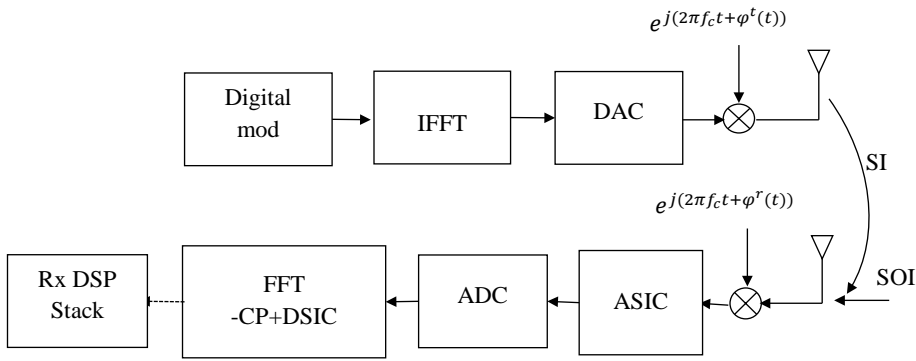

Fig. 1: Block Diagram of OFDM full-duplex transceiver

and *J*'s are the DFT of the transmitter and receiver PN and calculated as

$$J_k^i = \sum_{n=0}^{N-1} e^{j\phi_n^i} e^{-j2\pi nk/N} \quad (3)$$

Equation (2) can be simplified considering frequency-flat channel responses [8] to be as

$$Y_k = \sum_{l=0}^{N-1} X_l^I H_l^I \sum_{m=0}^{N-1} J_{m-l}^{t,I} J_{k-m}^r + Y_k^S + Z_k \\ = \sum_{l=0}^{N-1} X_l^I H_l^I J_{k-l}^c + Y_k^S + Z_k \quad (4)$$

where $J^c$ is the frequency-domain representation of the combined transmitter and receiver PN. Equation (2) can be rewritten as

$$Y_k = X_k^I H_k^I J_0^c + \sum_{l=0, l \neq k} X_l^I H_l^I J_{k-l}^c + Y_k^S + Z_k \quad (5)$$

where $J_0^c$ is called the common phase error (CPE) and the second term is the intercarrier interference (ICI) associated with the SI signal. The time domain representation of the previous equation can be written as

$$y_n = (x_n^I * h_n^I) j_n^c + y_n^S + z_n \quad (6)$$

where $j_n^c$ is the time domain representation of the combined PN which can be calculated as

$$j_n^c = e^{j(\phi_n^{t,I} + \phi_n^r)} \quad (7)$$

Equation (5) means that due to the presence of phase noise, each received subcarrier signal suffers from CPE and ICI components and is expressed as the weighted sum of all transmitted subcarrier signals multiplied by the corresponding phase error on each subcarrier. Moreover, both CPE and ICI are functions of those weighting coefficients, so that, once those weighting coefficients are obtained, phase noise can be mitigated [7].

Here, there are two transceiver implementations: the common versus independent oscillators. The choice of using a shared oscillator versus different oscillators for the transmitter and receiver of the same transceiver depends on the design requirements. Common oscillator is better chosen in small devices. However, in some setups where transmitter and receiver must be separated by a required distance, the choice of independent oscillator is mandatory.

Suppressing only the CPE can be viewed as the cancellation of the common rotation that equally affects all the subcarriers in the OFDM symbol or the DC level of the instantaneous phase noise. ICI expresses the effect of non-zero components of other subcarriers on the signal in certain subcarrier position due to the loss of orthogonality and the phase noise at that subcarrier due to the subcarrier-wise phase spread [7]. Some researchers introduced PN suppression schemes in the frequency domain after performing discrete Fourier transform (DFT)[5,8].

The choice of an efficient optimization algorithm with a maximum cancellable self-interference power and minimum computational complexity is the factor that determine the quality of such a SIC scheme. Other recent publications were interested in time-domain approach [4,6,8,9]. Time-domain PN suppression has the advantage of simplicity if compared with frequency domain as the phase noise is a multiplicative process in the time-domain while it is a convolution process in frequency domain. Another advantage of the time-domain approach is that it does not need OFDM-symbol synchronization [6] and is of higher accuracy than frequency-domain approach because the effect of PN is directly multiplied by the time sample but in frequency domain, it is distributed over all the used subcarriers and that complicates the exact solution of the optimization problem.

### III. THE PROPOSED TIME-DOMAIN PN ESTIMATOR

Although a robust phase noise estimation in full-duplex systems is crucial for better SIC performance, it is much easier than that for half-duplex systems, thanks to the total knowledge of the self-interference signal all the time along with the reception of the SOI data. This means that there is no need for blind estimators or the insertion of pilot samples. The phase noise estimation problem in FD can be formulated as an unknown system with a traceable statistically-random response.

Another factor that simplifies the phase noise estimation process is that the oscillator phase noise values are temporally-correlated and the PN is considered an almost stationary process over an optimum number of consecutive samples when modeled as Wiener process when using a free-running oscillator or as Ornstein-Uhlenbeck process in case of phase-locked-loop (PLL) based oscillators [14]. This leads to the block-wise processing of the phase noise estimation or, in other words, applying the PN estimate in a window-fashion manner with an optimized window size.

Referring to Fig. 2, the suggested implementation of the proposed time-domain WF-based estimator in the receiver side of the FD transceiver is shown. TD WF-based approach deals with the phase noise as an unknown response with its input and output are totally known. Here, TD WF-based estimator has the ability to discard any uncorrelated or unexpected change in its input-output relation. Hence, it can estimate the combined phase noise effect on the SI signal and reduce the level of SOI as a small uncorrelated and interfering noise through the PN estimation stage. Because the SI power is much stronger than the SOI power level, we can consider the SOI signal and the error in the SI channel estimation as noise through the PN estimation stage [6]. Hence, Equation (6) can be rewritten as

$$y_n = (x_n^I * h_n^I) j_n^c + noise \tag{8}$$

Assuming that the estimation of the interference channel was performed in an earlier stage and the input of the TD WF, $u_n$, is expressed as follows:

$$u_n = x_n^I * \hat{h}_n^I \tag{9}$$

The instantaneous estimation error in the value of SI sample n is

$$\begin{aligned} e_{yn} &= y_n - \hat{y}_n \\ &= y_n - u_n \cdot w_n \end{aligned} \tag{10}$$

and hence the optimal weight is expressed as

$$w_n^{opt} = \frac{\sum_{m=1+MK}^{M(1+K)} y_m \cdot u_m^*}{\sum_{m=1+MK}^{M(1+K)} u_m \cdot u_m^*} \tag{11}$$

where $K = n \bmod M$. This equation is the original WF design equation that deals with cross- and auto-correlation of the input and output samples within an optimal window size as different input-output (I/O) observations and averages its weights and then obtain the PN estimate from the averaged weight value and then applies this estimated value to all the input samples in a window-fashion manner [16]. In fact, this reduces the SOI level and any system impairments in addition to the channel estimation error as abrupt variations or an uncorrelated noise.

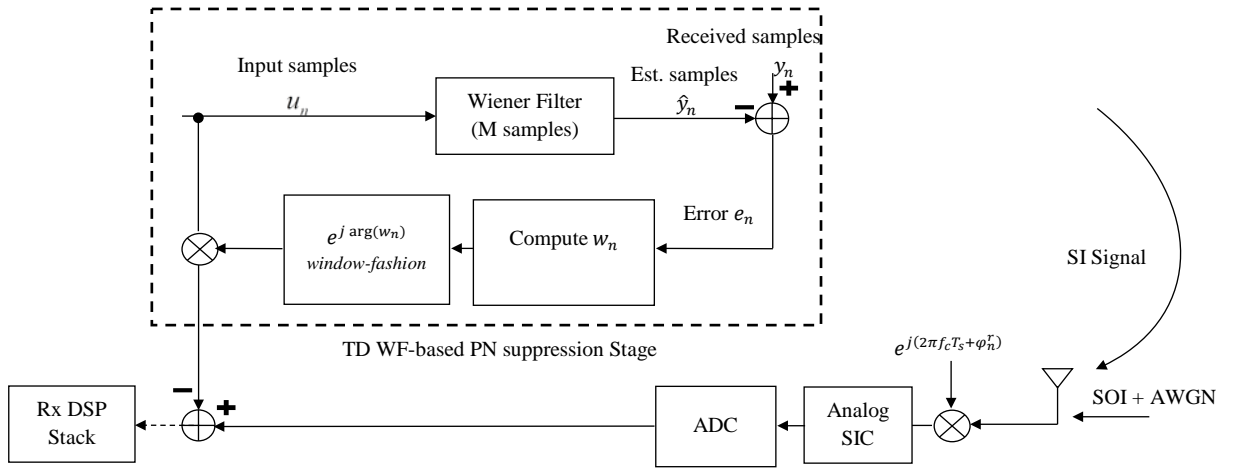

Fig. 2: The block diagram of the FD receiver with the proposed TD WF-based PN estimator

This highlights the advantage of using Wiener filtering instead of least squares (LS) based estimator in [15] for ICI estimation that may cause a negative gain problem [8]. The estimate of the instantaneous phase noise is then given by

$$\hat{\varphi}_n = \arg\left\{ \frac{\sum_{m=1+MK}^{M(1+K)} y_m \cdot u_m^*}{\sum_{m=1+MK}^{M(1+K)} u_m \cdot u_m^*} \right\} \quad (12)$$

For the phase noise mitigation stage, it is performed as a last single step by multiplying the incoming input signal by $e^{j\hat{\varphi}_n}$. The optimal window size is chosen to minimize the mean squared error (MSE) of the obtained phase value and this leads to that the optimal window size depends on the difference in the SOI to SI power levels and the variance of the phase noise process.

## IV. SIMULATION RESULTS AND ANALYSIS

Through this simulation, the common free-running oscillator approach is considered with the SI transmitter and receiver identical Wiener process PN. Wiener process or Brownian motion is modeled as

$$\varphi((n+1)T_s) = \varphi(nT_s) + N(0, 2\pi\beta T_s) \quad (14)$$

where $T_s$ is the sampling time and $\beta$ is the 3-dB bandwidth of the phase noise Lorentzian spectrum and is related to the quality factor of the used oscillator. This relation means that the transition phase noise value at $(n+1)T_s$ from its value at $nT_s$ is described by a Gaussian random variable (RV) with 0 mean and $2\pi\beta T_s$. From (14), Wiener process is considered an almost-stationary process at low 3dB-BW values.

In this section, the proposed TD WF-based PN estimator is compared to the well-known 'Only-CPE' PN estimation algorithm and a recent time-domain PN estimator presented in [6] based on the low pass filtering (LPF) which has a better performance than the frequency-domain estimator in [8] by 9 dB with a lower computational complexity.

The used numerology through this simulation is chosen according to the LTE-downlink theme. A transmission frame of 64 OFDM symbols with 1024 subcarriers of which 300 subcarriers are dedicated to the user data each carries a 16-QAM data symbol at a sampling frequency of 15.36 MHz. A Rician fading channel with a Gaussian channel estimation error is assumed and the antenna separation and analog SIC are set to 30 dB for each, as those described in [6].

Fig. 3 shows the achieved digital suppression of the residual SI power versus the oscillator phase noise 3-dB bandwidth after performing proper antenna separation and analog SIC with a fixed value of SIR = -30 dB. It is noticed that the proposed algorithm has a superior performance than the PN estimator based on LPF especially in high values of PN within the realistic full-duplex range. It can achieve more than 4 dB enhancement with a noticeable reduction in the computational complexity as will be discussed later.

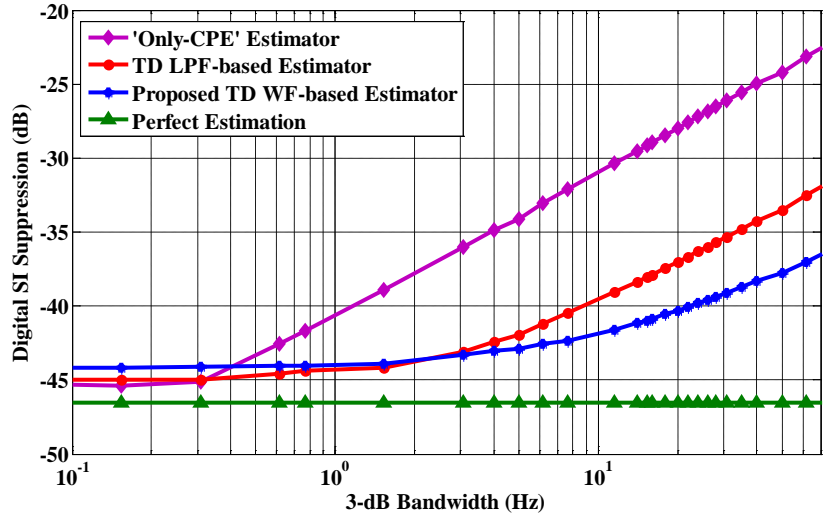

Fig. 3: The achieved digital SI suppression of the proposed algorithm, Only-CPE estimator and the LPF PN estimator versus PN 3dB BW

Fig. 4 describes the relation between the maximum achievable SI suppression and the SOI to SI average power levels after the antenna separation and channel defined as the channel attenuation difference at a fixed value of 3-dB PN bandwidth of 10 Hz. This figure, in fact, describes the impact of the SOI presence on the phase noise estimation stage. It is clear that the performance of the proposed algorithm and the LPF-based algorithm are lower than the performance of using 'Only-CPE' estimation below -42 to -45 dB channel attenuation difference before the analog SIC which is not realistic for full-duplexing because in high SOI to SI power level ratio, a part of SOI power is dominant and existent along with the SI signal that makes the WF unable to classify it as abrupt or unexpected variations in its I/O relation.

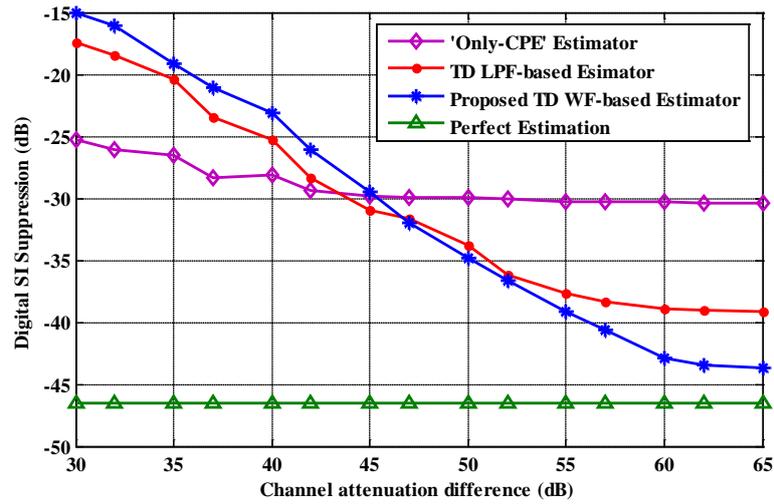

Fig. 4: The achieved digital SI suppression versus the difference of the SOI power level and SI power level before analog SIC

The used window size is numerically optimized in Fig. 5. The used window size is related to the value of the SOI to SI power levels as the SOI is considered a noise in the estimation stage. It is also related to the PN linewidth which is defined as the product of the 3-dB bandwidth $\beta$ and the symbol duration $T_s$ which describes the variance of the phase noise as Wiener process. Small window size makes the proposed algorithm performance tendency towards the LPF-based algorithm performance as it loses the temporal correlation of the Wiener process. While, large window size pushes its performance to the 'Only-CPE' estimator as the averaging process over large OFDM symbol size increases the instanteous estimation error especially with high values of PN bandwidth.

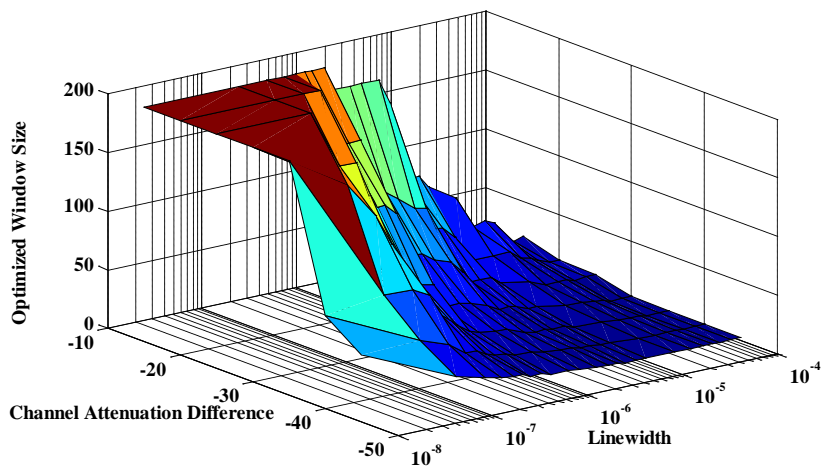

Fig. 5: The averaging window size optimization at different ratios of SOI to SI levels

The computational complexity of the proposed algorithm is compared to the time-domain approaches presented in [6] and [8] in Table I. where $N$ is the number of the estimated samples / OFDM symbol in [8] and it was optimized to be $K/2$ where $K$ is the total number of samples/OFDM symbol, $M$ is the optimized window size of the proposed algorithm=35 samples, and $L$ is the used filter length in [6] and it was optimized to be $L$=50. It was investigated that the complexity of the time-domain approach, even when the complexity of IDFT$\{X_k.H_k\}$ is added, is lower than the frequency-domain one. In the phase-noise mitigation part, the time-domain algorithm require single complex multiplication per sample for the proposed algorithm and LPF-based estimator and an additional arithmetic operation per sample for interpolation, whereas frequency-domain approaches require convolving the useful signal with the signal whose length is the same as the amount of estimated phase-noise spectral components.

These simulation results shows a very good differentiator between the proposed algorithm and the already-existing estimators. With low values of channel attenuation differences (i.e. high SIR values) and small PN variance, the proposed algorithm tends to act like 'Only-CPE' estimator while it has the ability of the LPF-based estimator to track higher PN variance with a relatively high values of channel attenuation difference. This point of view, puts the proposed algorithm as an intermediate solution between the commonly used 'Only-CPE' estimator and the LPF-based estimator with a clearly reduced computational complexity and superior performance.

TABLE I: THE COMPUTATIONAL COMPLEXITY OF DIFFERENT TD PN-ESTIMATORS

|  | Proposed estimator | TD LPF-based estimator | TD MMSE-based estimator |
|---|---|---|---|
| Complexity | $O(4M+1)$ | $O(3L)$ | $O(N^2)$ |
| Arithmetic Operations/Sa | ≈4 | 150 | 256 |

## V. CONCLUSION

In full-duplex systems, phase noise mitigation is considered a crucial step for SI cancellation. It was found that time-domain estimators are considered the better candidate for that than frequency-domain estimators thanks to their good performance with lower complexity. This paper presents a novel time-domain phase noise estimator based on the idea of Wiener filtering with a proper averaging-window size. The proposed algorithm shows superior SI suppression results when compared to 'Only-CPE' estimator and time-domain LPF-based estimator with a noticeable reduction in the required computational complexity.